\documentclass[manuscript]{aastex}

\newcommand{\kep}{\textit{Kepler} }
\newcommand{\ms}{m~s$^{-1}$}

\shorttitle{Absolute parameters of KIC~09246715}
\shortauthors{He\l miniak et al.}

\begin{document}


\title{Absolute stellar parameters of KIC~09246715 -- a double-giant\\
eclipsing system with a solar-like oscillator
}


\author{K. G. He\l miniak\altaffilmark{1}}
\affil{Subaru Telescope, National Astronomical Observatory of Japan,
\\650 N. Aohoku Place, Hilo, HI 96720, USA}
\email{xysiek@naoj.org}

\and

\author{N. Ukita\altaffilmark{2} and E. Kambe}
\affil{Okayama Astrophysical Observatory, National Astronomical Observatory of Japan,
\\3037-5 Honjo, Kamogata, Asakuchi, Okayama 719-0232, Japan}

\and

\author{M. Konacki}
\affil{Nicolaus Copernicus Astronomical Center, Department of Astrophysics,
\\ul. Rabia\'nska 8, 87-100 Toru\'n, Poland}


\altaffiltext{1}{Subaru Research Fellow}
\altaffiltext{2}{Graduate University for Advanced Studies, 2-21-1 Osawa, Mitaka, Tokyo 181-8588, Japan}


\begin{abstract}
We present our results of a combined analysis of radial velocity and light curves
of a double-lined spectroscopic and eclipsing binary KIC~09246715, observed
photometrically by the \kep satellite, and spectroscopically with the OAO-1.88m
telescope with the HIgh-Dispertion Echelle Spectrograph (HIDES). The target was 
claimed to be composed of two red giants, one of which is showing solar-like
oscillations. We have found that the mass and radius of the primary are 
$M_1=2.169\pm0.024$~M$_\odot$ and $R_1=8.47\pm0.13$~R$_\odot$, and of the secondary:
$M_2=2.143\pm0.025$~M$_\odot$ and $R_2=8.18\pm0.09$~R$_\odot$, which confirms the
double-giant status. Our secondary is the star to which the oscillations 
were attributed. Results of its previous asteroseismic analysis
are in agreement with ours, only significantly less precise, but the subsequent
light-curve-based study failed to derive correct mass and radius of our primary.
KIC~09246715 is one of the rare cases where asteroseismic parameters
of a solar-like oscillator were confirmed by an independent method, and only the
third example of a Galactic double-giant eclipsing binary with masses and radii
measured with precision below 2\%.
\end{abstract}





\section{Introduction}
The unprecedented photometric precision of the \kep mission \citep{bor10}
has opened new possibilities in various fields of astronomy, including studies 
of eclipsing binaries and asteroseismology. The former has been known for many decades 
to provide accurate and precise stellar parameters, sometimes not possible to 
derive with other methods (e.g. absolute radii). Since the discovery of 
pulsating sdB stars \citep{kil94} and solar-like oscillations in stars other 
than the Sun \citep{bro94,kje95}, asteroseismology
became another technique capable of providing such results. Until the 
launch of {\it CoRoT} and \kep satellites the number of other stars showing
the solar-like oscillations was very low, but now it is counted in thousands,
and contains main sequence, as well as giant stars.

Absolute values of fundamental stellar parameters have multiple 
applications in modern astronomy, but they need to be known with precision 
of at least 2--3\% \citep{las02}. The best possible characterisation
comes from double-lined spectroscopic and eclipsing binaries, that also
show pulsations. Such systems are extremely useful for testing stellar evolution
models, as the masses and radii can be, in principle, measured with two 
independent techniques. Around 50 eclipsing systems containing oscillating giants
have been reported by \citet[][hereafter: G13]{gau13}, but without spectroscopic follow up.
Only \object{KIC~08410637}, composed of an oscillating giant
and a main-sequence star, has been studied in more detail \citep{fra13},
and a very good precision in absolute parameters has been achieved.
Despite being very useful in general, currently the asteroseismic measurements 
usually do not give results of sufficient precision. Nevertheless, the large number 
of known oscillating stars undoubtedly shows the importance of this technique.
Discovery and characterisation of the ``keystone'' double-lined spectroscopic,
eclipsing and oscillating systems will help to test and enhance the capabilities of 
asteroseismology directly. In this work we present such an example.

\section{The target}

The system \object{KIC~09246715} (HD~190585, $V=9.65$, hereafter: K0924)
was identified as an eclipsing binary by the \kep mission, and was listed in the 
\kep Eclipsing Binaries Catalog \citep[KEBC;][]{prs11,sla11}. It was not reported
as a double-lined spectroscopic binary till now. According to {G13} and 
\citet[][hereafter: G14]{gau14},
it is composed of two red giants, one of which is showing solar-like oscillations
with peak frequency and separation
of $\nu_{max}=102.2$~$\mu$Hz, $\Delta\nu=8.3$~$\mu$Hz ({G13}), 
or $\nu_{max}=106.38(75)$~$\mu$Hz, $\Delta\nu=8.327(10)$~$\mu$Hz ({G14}). 
Activity is also reported, in a form of periodic ($P_{var}=93.3$~d) brightness 
modulations coming from spots.

Using asteroseismology G13 and G14 first derived mass and radius of the
oscillating star. Then they combined it with results of light curve modelling
(i.e. $R_1/a,R_2/a$) and Kepler's 3rd law, inferring the properties 
of the other component: 2.06(13)~M$_\odot$, 8.10(18)~R$_\odot$ for the oscillating 
component (their primary), and 1.1(3)~M$_\odot$, 6.8(1)~R$_\odot$ or 3.3(5)~M$_\odot$, 
9.7(2)~R$_\odot$ for the other, but described the latter solution as less probable.
With our spectroscopy, we are able to directly revise their results,
confronting the asteroseismology with direct modelling of the light and 
radial velocity (RV) curves of this system.

Eclipsing binaries composed of two red giants, like K0924, are
relatively rare, and not many such systems have their 
components parameters measured with high precision. The on-line DEBCat catalog
\citep{sou15}, which lists detached eclipsing binaries with masses and radii
measured down to 2-3\%, contains only two such Galactic systems -- 
HD~187669 \citep{hel15} and ASAS~J180052-2333.8 \citep{suc15} -- and a handful of 
others found in LMC and SMC \citep[e.g.][]{pie13,gra14}. Therefore K0924 is also 
important for studying the late stages of stellar life and testing the models of evolution 
and structure.

\section{Data and methodology}
The system K0924 is one of the targets we have observed as a part of a larger
spectroscopic project, aimed for characterisation of northern detached 
eclipsing binaries, including
objects selected from the \kep field. All of them have been observed and analysed 
in a similar way, and the detailed description of the observing procedure, data
reduction, stability tests, orbit and light curve fitting, etc. will be explained
in a forthcoming paper (He\l miniak et al., in prep.). However, we shortly 
present them below.

We follow the convention that the primary component is the one being eclipsed
during the deeper minimum (the primary eclipse). In our case, 
the primary is the slightly more massive and larger star.

\subsection{Observational data}
In this study we make use of the publicly available \kep mission photometry.
We used the de-trended relative flux measurements $f_{dtr}$, 
that were later transformed into magnitude difference $\Delta m=-2.5\log(f_{dtr})$,
and finally the KEBC value of $k_{mag}$ was added. 
The de-trended data were downloaded directly from KEBC, where currently only the
long-cadence measurements for this star are available.

A total of 8 spectra were taken during several runs between July 2014
and late April 2015, at the 1.88-m telescope of the Okayama Astrophysical Observatory 
with the HIgh-Dispersion Echelle Spectrograph \citep[HIDES;][]{izu99}. The instrument 
was fed through a circular fiber, for which the light is collected via a circular 
aperture of projected on-sky diameter of 2.7~arcsec, drilled in a flat mirror 
that is used for guiding \citep{kam13}. An image slicer is used in order to reach 
both high resolution ($R\sim50000$) and good efficiency of the system.
Wavelength calibration was based on ThAr lamp exposures taken every 1-2 hours,
which gave the stability of the instrument at the level of 40-50 \ms, as measured
from multiple observations of four RV standards. Each science exposure was 1500 seconds 
long, but due to unstable atmospheric conditions, the resulting signal-to-noise
ratio ($SNR$) usually varied between 75 and 105 (63 at one time). Spectra were reduced 
with dedicated IRAF-based scripts, that cope with the mosaic character of the HIDES detector. 
Reduction included correction for overscan, bias, flat field, cosmic rays and bad pixels,
as well as extraction and wavelength calibration of 53 orders (4360--7535 \AA).

\subsection{Light curve analysis}

For the light curve (LC) fit we used version 28 (v28) 
of the code JKTEBOP \citep{sou04a,sou04b}, which is based on the 
EBOP program \citep{pop81}. The complete long-cadence Q0-Q17 curve was 
first used to find the best-fitting parameters, including period $P$, 
primary (deeper) eclipse mid-time $T_0$, eccentricity $e$, periastron longitude 
$\omega$, inclination $i$, ratio of fluxes $L_2/L_1$, ratio of central surface 
brightnesses $J$, sum of the fractional radii $r_1+r_2$ (in units of
major semi-axis $a$), and their ratio $k$. A small but clear increase in brightness was 
noticed around both eclipses, so we also fitted 
reflection coefficients, and got $7.4(1.0)\times10^{-4}$ and $6.3(1.0)\times10^{-4}$.
Logarithmic limb darkening (LD) law \citep{kin70} was assumed, with coefficients
(fixed in the fit) interpolated from the the tables published on the PHOEBE
website\footnote{\tt http://phoebe-project.org/1.0/?q=node/110}. For this, 
the gravities $\log(g)$ were taken from an initial fit (corrected later), 
which results, i.e. masses and radii, were then compared to the PARSEC 
isochrones \citep{bre12} in order to estimate the temperatures.
The gravity darkening coefficients were always kept fixed at the values 
appropriate for stars with convective envelopes ($g=0.32$).
At the end, the code calculates the fractional radii $r_{1,2}$, and 
fluxes $L_{1,2}$. As input we also used spectroscopic flux ratios as found
by TODCOR (see Sec. \ref{sec_rv}).

The JKTEBOP does not fit for spots or 
oscillations, however, it offers a number of algorithms to properly include 
systematics in the error budget. We assumed that after years of nearly continuous 
observations they are averaged out over the orbital period, and decided to treat 
them as a correlated noise and run the residual-shifts (RS) procedure to calculate 
reliable uncertainties \citep{sou11}. To run RS on the whole Q0-Q17 curve would 
take almost two weeks on the computer we used, so for the errors estimation we 
decided to split the data and analyse each chunk separately, and then calculate weighted 
averages of the resulting parameters. Due to its long period, reaching almost the 
time span of two quarters of \kep data, both eclipses were not always visible during 
the same quarter. Adjacent quarters with only primary and only secondary minimum
were merged together into one set and cropped, so the resulting light curve
time span was around 90 days and covered both minima. This was done for 
Q6+Q7, Q8+Q9, Q10+Q11, and Q12+Q13. Both eclipses were observed in Q3, Q5 and Q14.
No eclipses were seen in Q0, Q1, Q2, Q4, Q15, Q16 and Q17, which were 
rejected from the RS stage. Because of the long period, $P,T_0,e,\omega$ and 
reflection coefficients, as well as the LD coefficients, were 
first held fixed, only perturbed later during the proper RS stage.
The final parameter errors were obtained by adding in quadrature the
formal error of the weighted average and the {\it rms} of the results
for each quarter.

\subsection{Radial velocities and orbital fit}\label{sec_rv}
RVs were measured with our own implementation of the two-spectra cross-correlation 
function technique TODCOR \citep{zuc94} with synthetic spectra computed with 
ATLAS9 and ATLAS12 codes \citep{kur92} as templates. Single measurement 
errors were calculated with a bootstrap approach \citep{hel12}. Flux ratios
\citep[$\alpha$ in][]{zuc94} were also calculated, and used as input for
JKTEBOP. For this, only orders from the \kep bandpass were used, with weights corresponding
to the total response factor at a wavelength equal the center of a given
spectral order. The RVs obtained from our HIDES spectra, together with their 
errors and $SNR$ of the spectra, are listed in Table~\ref{tab_rv}.
\placetable{tab_rv}

The RV measurements were analysed with the code V2FIT, which fits a 
double-Keplerian orbit by using the Levenberg-Marquartd algorithm. As free 
parameters, we set the time of periastron passage $T_p$, the velocity 
semi-amplitudes $K_{1,2}$, the primary's systemic velocity $\gamma_1$, and 
the difference in components systemic velocities $\gamma_2-\gamma_1$.
Period $P$, eccentricity $e$ and periastron longitude $\omega$ were kept
fixed on values found in JKTEBOP. In order to find reliable uncertainties, that
include the influence of systematics coming possibly from the pulsations
and low number of spectra, we run 10000 bootstrap iterations. As expected, we
found the systematics to be the dominant source of parameter errors.

\subsection{Absolute values of parameters}
The absolute values of parameters and their uncertainties were calculated 
with the JKTABSDIM code, available together with JKTEBOP. This simple code 
combines the spectroscopic and light curve solutions to derive a set of 
stellar absolute dimensions ($M_{1,2},R_{1,2},a$), and related quantities
($L_{1,2}/L_\odot,M_{bol1,2},\log(g_{1,2})$). If desired, it also calculates
distance, but requires multi-color photometry, $E(B-V)$, and both effective
temperatures as input. Due to lack of such data, we did not attempt to
estimate the distance.

\section{Results}

\begin{figure}
\epsscale{.90}
\plotone{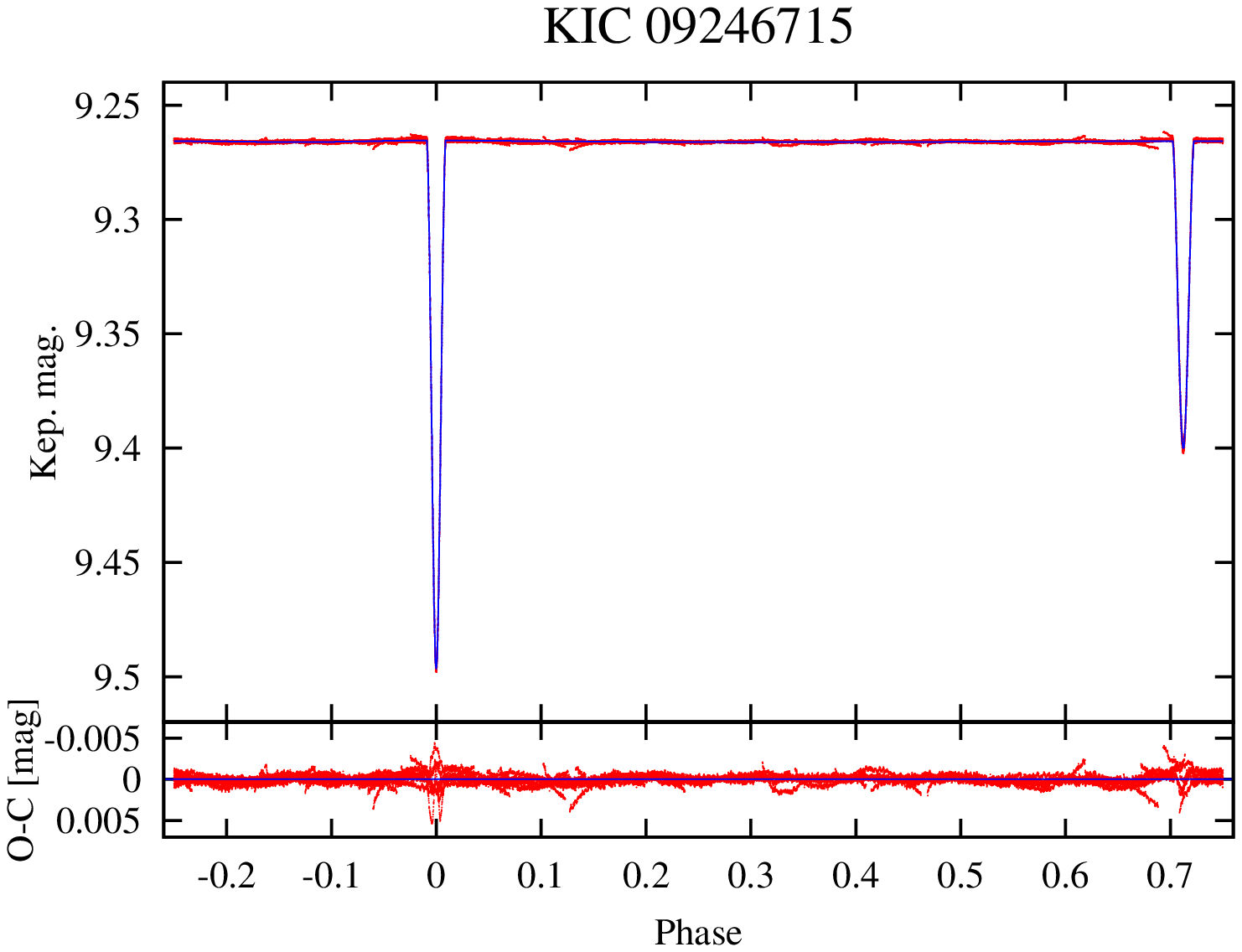}
\caption{The observed (red) and model (blue) Q0-Q17 light curves of K0924.
Phase 0 is for the deeper eclipse mid-time. Residuals are shown in the
lower panel.}\label{fig_lc_0924}
\end{figure}

\begin{figure}
\epsscale{.90}
\plotone{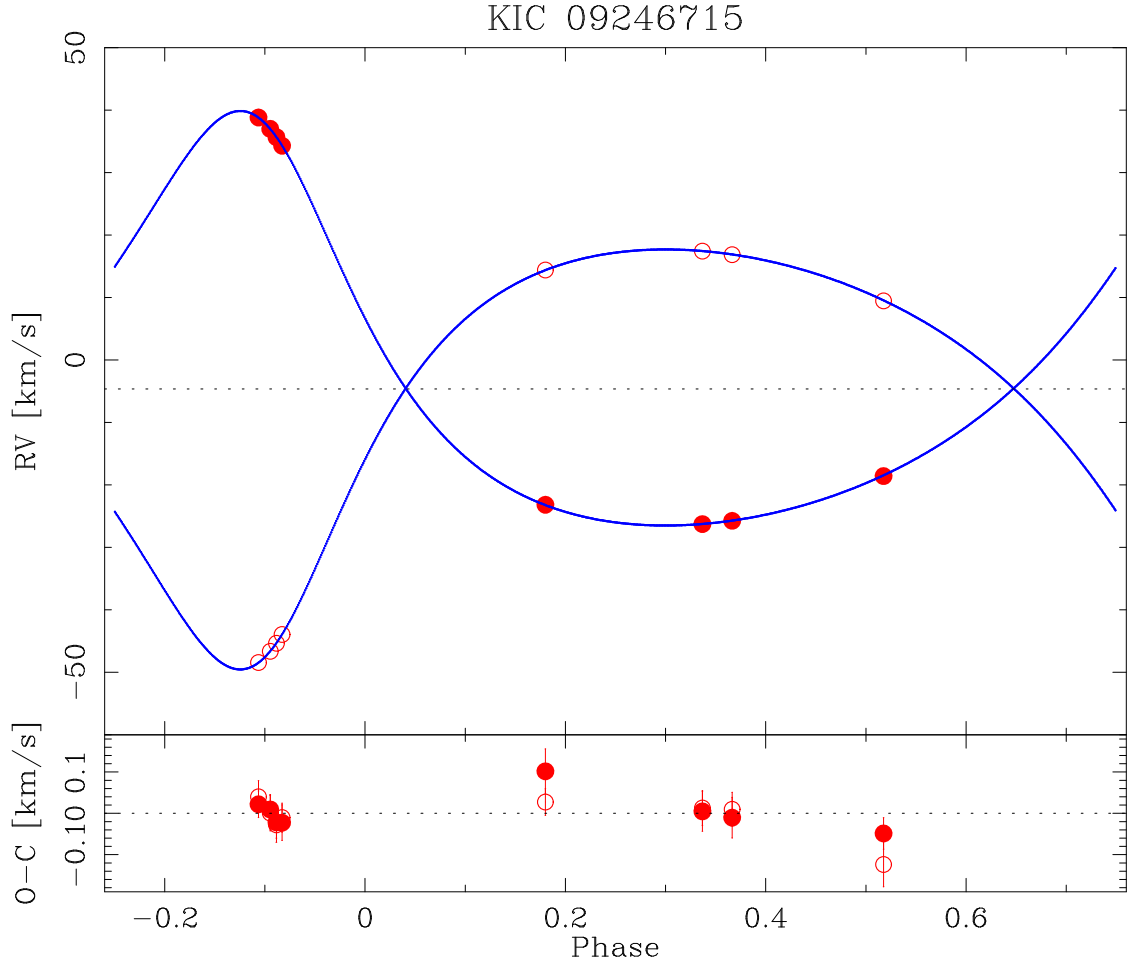}
\caption{Radial velocity curves and of KIC~09246715. 
Filled red circles refer to the primary, and open ones to the secondary.
The blue lines are the best-fitting model curves. Systemic velocity is
marked by the dotted line. Residuals are shown in the lower panel. Ephemeris
are the same as in Fig. \ref{fig_lc_0924}.
}\label{fig_rv_0924}
\end{figure}

The light and RV curves are presented in Figures 
\ref{fig_lc_0924} and \ref{fig_rv_0924}, respectively. 
Values of stellar parameters can be found in Table \ref{tab_par}.
We have found that the mass and radius of the primary are 
$M_1=2.169\pm0.024$~M$_\odot$ and $R_1=8.47\pm0.13$~R$_\odot$, and of the secondary:
$M_2=2.143\pm0.025$~M$_\odot$ and $R_2=8.18\pm0.09$~R$_\odot$.
K0924 is composed of two very similar stars that evolved significantly 
from the main sequence. We reached a good precision of $\sim$1.1\%
in masses and 1.1--1.5\% in radii, which makes our results valuable for further 
comparison with models of stellar structure and evolution, and places 
K0924 in a very small group of double-giant systems with precisely
measured masses and radii. Precision in masses is mainly hampered by the
low number of spectra, and likely by the influence of spots, oscillations, and 
instrument distortions on the RV measurements, which may explain the apparently 
non-random residuals. Precision in radii is lowered by the oscillations, which
we did not model and reduce, but treated as a red noise in the light curve, and 
included in the error budget.

\placetable{tab_par}
\placefigure{fig_lc_0924}
\placefigure{fig_rv_0924}

These results are different from G13, who give 1.7(3)~M$_\odot$ 
and 7.7(4)~R$_\odot$ for the oscillating component, and 0.8(7)~M$_\odot$ and 
5.9(3)~R$_\odot$ for the other, as well as from G14, who give, analogously, 
2.06(13)~M$_\odot+8.10(18)$~R$_\odot$, and 1.1(3)~M$_\odot+6.8(1)$~R$_\odot$.
Note that both previous studies adopt the opposite definition of primary/secondary, 
and solar-like oscillations are found on the star that we define as the secondary, 
i.e. slightly less massive according to our orbital solution. First, both 
stars have almost equal masses and radii, while the solution of G14 
gives $M_2/M_1\simeq1.87$ and $k\simeq1.19$. Our mass ratio close to 1 
comes directly from the spectroscopy (Fig. \ref{fig_rv_0924}), and its value is 
very robust. The ratio of radii was constrained using the spectroscopic flux ratios from 
TODCOR, which we also find reliable -- a quick inspection of any of 
the spectra reveals that the lines of both components are of similar depths, 
meaning similar effective temperatures and levels of continuum, and the
peaks of cross-correlation function were of basically the same height. 
Finally, a quick comparison with evolutionary models (see next Section) 
shows that there is no isochrone that reproduces the results of previous 
studies. This means that the asteroseismology 
provides correct, yet less precise, stellar parameters, but light curve 
solutions should be supported by spectroscopic values of mass and flux ratios.
It has also been found in the case of KIC~08410637 by \citet{fra13}.
Moreover, some conclusions presented in G14 should now be treated with
caution.

It is worth to note that despite very large separation, the system might have 
reached or is close to a form of a tidal equilibrium. If the reported period 
$P_{var}=93.3$~d is related to the rotation, it would coincide with the 
value of $P_{rot,ps}=94.2(3)$~d, which is the value of rotation expected for the 
given orbital period and eccentricity in the state of pseudo-synchronisation
\citep{hut81,maz08}.

\section{Discussion}
\begin{figure}
\epsscale{.90}
\plotone{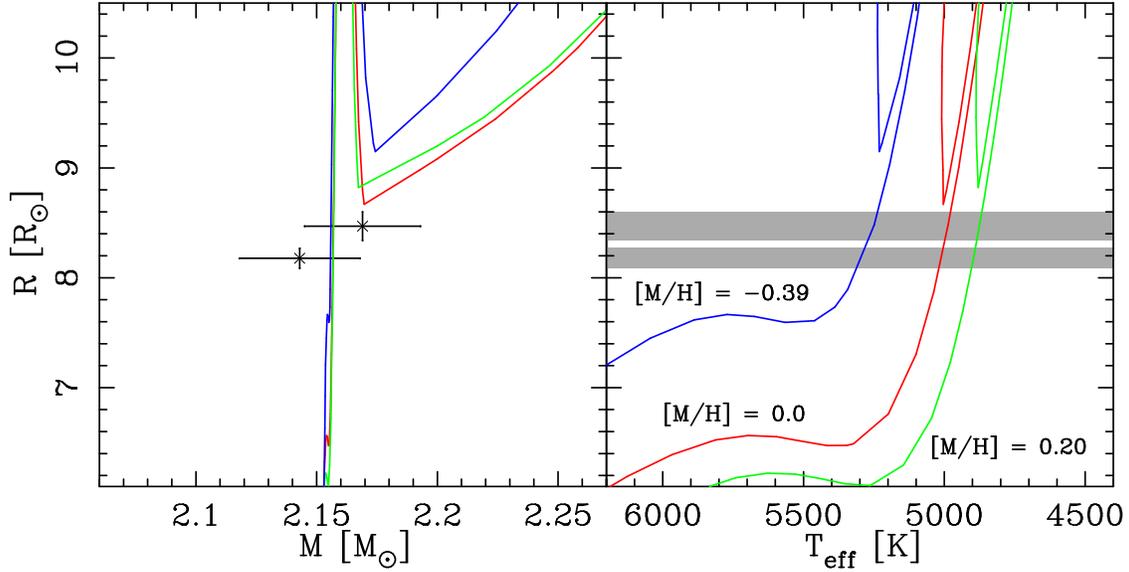}
\caption{{\it Left:} Comparison of our results with PARSEC isochrones
for $[M/H]=0.0$ and age 950 Myr (red), $[M/H]=-0.39$ and age of 
780~Myr (blue), and $[M/H]=0.20$ and age 1.04 Gyr (green) on the mass-radius 
plane. For a given metallicity, the age is
restricted mainly by the precision in masses. {\it Right:} Same isochrones
on the temperature-radius plane. Grey stripes mark the $1\sigma$ ranges of
radii we obtained in our analysis. High precision in $R$ allows us to
estimate temperatures expected for each metallicity/age.
With independent $[M/H]$ or $T_{eff}$ measurement, coming from spectral 
analysis for example, the age-metallicity degeneracy would be solved.
}\label{fig_iso}
\end{figure}

In Figure \ref{fig_iso} we compare our results, i.e. masses and radii, 
with the theoretical PARSEC isochrones \citep{bre12}, that include 
calculation of absolute magnitudes in the \kep photometric band. 
In this set of models the solar metallicity is reached for $Z$=0.0152.

\placefigure{fig_iso}

From Figure \ref{fig_iso} one can deduce that the system may be $\sim$950~Myr old, 
if solar metallicity is assumed. At this stage, however, the age and $[M/H]$
are strongly degenerated on the mass-radius plane \citep{hel15}. 
Both components are probably on the red giant branch, before the red clump stage.
It would be possible to reproduce both the observed radii on the red clump, by assuming 
a lower metallicity ($[M/H]<-1.6$), but then the system would be even younger 
(600--700~Myr) and of much earlier spectral type than suggested by the KEBC
effective temperature of 4699~K. Both components likely have nearly equal
$T_{eff}$'s, and the spectra show many features, so we find the $\sim$4700~K,
$\sim$950~Myr and solar metallicity solution more probable, however with some
caution about the temperature itself (see further text). 

We also checked the value of metallicity of $-0.39$~dex, given in the Mikulski 
Archive for Space Telescopes (MAST)\footnote{\tt http://archive.stsci.edu/index.html},
which was, however, obtained from analysis of combined light of both stars. 
It is listed together with $\log(g)=2.42$, which is not in agreement with 
our results, so we treat this value of $[M/H]$ with caution, but we cannot exclude 
it completely. The best-fitting PARSEC isochrone for this $[M/H]$ is for the age of 
780~Myr, and is also shown in Fig. \ref{fig_iso}.
For completeness, we also present an isochrone for a more metal abundant case
$[M/H]$=0.20. The best-fitting age is then 1.04 Gyr

Due to rapid changes in the stellar structure during the red giant phase (compared 
to the main sequence), the isochrones predict a variety of possible temperatures 
within the range of masses we found for K0924. However, the evolution in
$T_{eff}$ is related to the growth of the star, so we can use the radii
estimates to predict temperatures of both components. However, due to lack 
of any $[M/H]$ estimate that we find reliable, we 
can not pick any certain scenario. 
On the 950~Myr, solar-metallicity isochrone, we find $T_{eff,1}=4990(11)$~K, and 
$T_{eff,2}=5013(7)$~K. On the 780~Myr, $-0.39$~dex isochrone we get
$T_{eff,1}=5250(20)$~K, and $T_{eff,2}=5300(12)$~K. Finally, the 1.04~Gyr, 0.20~dex
gives $T_{eff,1}=4876(9)$~K, and $T_{eff,2}=4896(7)$~K. To obtain temperatures 
around 4700~K, one would have to use an isochrone for even higher metallicity. 
The difference in predicted temperatures is large 
enough for the modern spectral analysis methods to distinguish between
solar and sub-solar cases. Note also that the ratio of
predicted values of temperatures is much closer to 1 than 1.04, which was found by 
\citep{gau14} in the light-curve fit, and that in their solution the oscillator
(our secondary) is cooler. The K0924 system would benefit from further 
spectroscopy (to disentangle the component spectra) 
and multi-band photometry, that altogether would allow for precise temperature 
and metallicity determination. It would also improve the age estimation,
and precision in masses and radii even more. 

Finally, taking into account the similarities of the two components, we suspect 
that both stars may show similar oscillations that blend in the light curve, and may 
have confused {G14}, although the mass and radius of their pulsating
component, and the 2014 values of $e,i$ and $(R_1+R_2)/a$ are in agreement 
(within errors) with our solution. 
We used our results to reproduce the expected peak frequency $\nu_{max}$
and frequency splitting $\Delta \nu$, using the formulae of \citet{kje95}:
\begin{equation}
\nu_{max} = \frac{M/M_\odot}{(R/R_\odot)^2 \sqrt{T_{eff}/5777}}\,3104\,\mu \mathrm{Hz},
\label{eq_dnu}
\end{equation}
\begin{equation}
\Delta \nu = \left(\frac{M}{M_\odot}\right)^{1/2} \left(\frac{R}{R_\odot}\right)^{-3/2} 138.8\,\mu \mathrm{Hz},
\end{equation}
with the reference values from \citet{mos13}, also used in G13.
Our results predict $\Delta\nu=8.29(20)$ and 8.69(15)~$\mu$Hz for the primary and secondary
(oscillator in G13 and G14), respectively. The peak frequency is however 
dependent on the temperature, and for the discussed range of 4700--5300~K we get 
104--98~$\mu$Hz for the primary, and 110--104~$\mu$Hz for the secondary. In particular,
the peak frequencies differ by less than the separation $\Delta \nu$ and perhaps 
two modulations have been mixed in the analysis of G13 and G14.

If we take that only one oscillation is really present, we can find the 
secondary's temperature from the equation \ref{eq_dnu}. To get
the value of $\nu_{max}$ reported in G14, the secondary should have 
$T_{eff,2}=5040(260)$~K, so metallicity close to solar would be
preferred ($[M/H]>-0.39$ from Fig. \ref{fig_iso}).

\section{Summary}
We have obtained precise mass and radius measurements of two giant
components of a detached eclipsing binary KIC~09246715. The secondary component
was previously reported to show solar-like oscillations, and our results
confirm mass and radius derived from asteroseismology. We have, however,
found that the other star -- our primary -- has very similar properties,
thus we suspect that both components may in fact show very
similar oscillations. We may conclude that the asteroseismology is a promising 
method of measuring stellar masses and radii, and that the correct estimation 
of these parameters for both components of a binary can not be done from 
light curves only, but still requires spectroscopic information.



\acknowledgments
We would like to thank Prof. M\'{a}rcio Catelan from the Pontificia Universidad
Cat\'{o}lica and the anonymous Referee for the discussion and comments.
KGH acknowledges support provided by the National Astronomical Observatory 
of Japan as Subaru Astronomical Research Fellow.
MK is supported by a European Research Council Starting Grant, the Ministry of 
Science and Higher Education (grant W103/ERC/2011) and the Foundation for Polish 
Science through grant ``Ideas for Poland''.



{\it Facilities:} \facility{Kepler}, \facility{OAO-1.88m (HIDES)}

\clearpage



\clearpage

\begin{table}
\begin{center}
\caption{Measured radial velocities of K0924 with their errors and residuals
of the orbital fit (all in k\ms). $SNR$ at 5800 \AA\ for each observation is also given.
\label{tab_rv}}
\begin{tabular}{lrrrrrrr}
\tableline\tableline
BJD-2450000 & $RV_1$ & $\epsilon_1$ & $(O-C)_1$ & $RV_2$ & $\epsilon_2$ & $(O-C)_2$ & $SNR$\\
\tableline
 6865.077724 & 38.818 & 0.032 & 0.022 &-48.417 & 0.039 & 0.040 & 105\\
 6867.078865 & 36.997 & 0.036 & 0.009 &-46.627 & 0.043 & 0.001 & 75\\
 6868.138442 & 35.671 & 0.038 &-0.022 &-45.346 & 0.042 &-0.028 & 63\\
 6869.109798 & 34.299 & 0.043 &-0.023 &-43.941 & 0.035 &-0.010 & 90\\
 6914.138901 &-23.176 & 0.054 & 0.102 & 14.398 & 0.032 & 0.027 & 101\\
 6946.075069 &-25.749 & 0.049 &-0.010 & 16.872 & 0.041 & 0.010 & 103\\
 7112.264233 &-26.269 & 0.048 & 0.005 & 17.417 & 0.042 & 0.013 & 81\\
 7143.230688 &-18.592 & 0.038 &-0.049 &  9.454 & 0.054 &-0.124 & 86\\
\tableline
\end{tabular}
\end{center}
\end{table}

\clearpage

\begin{table}
\begin{center}
\caption{Absolute orbital and physical parameters of K0924. Errors include systematics.
\label{tab_par}}
\begin{tabular}{lrr}
\tableline\tableline
Parameter   & Value &  $\pm$ \\
\tableline
$P$ (d) 	& 171.2770  & 0.0006 \\
$T_0$ (JD-2454900) & 99.2536 & 0.0031 \\
$T_P$ (JD-2454900) & 81.853  & 0.058 \\
$K_1$ (k\ms) &  33.18 & 0.16 \\
$K_2$ (k\ms) &  33.58 & 0.14 \\
$\gamma_1$ (k\ms)& -4.643 & 0.071 \\
$\gamma_2-\gamma_1$ (k\ms)& 0.15 & 0.12 \\
$q$		& 0.9880 & 0.0063 \\
$e$		& 0.3587 & 0.0009 \\
$\omega$ ($^\circ$) & 19.84 & 0.44 \\
$r_1$		& 0.04008 & 0.00058 \\
$r_2$		& 0.03870 & 0.00040 \\
$i$ ($^\circ$)	& 87.049 & 0.031 \\
$J$		    & 1.042 & 0.049 \\
$L_2/L_1$	& 0.964 & 0.048 \\
$M_1$ (M$_\odot$)& 2.169 & 0.024 \\
$M_2$ (M$_\odot$)& 2.143 & 0.025 \\
$R_1$ (R$_\odot$)&  8.47 & 0.13 \\
$R_2$ (R$_\odot$)&  8.18 & 0.09 \\
$a$   (R$_\odot$)& 211.29& 0.77 \\
$\log(g_1)$	 & 2.919 & 0.013 \\
$\log(g_2)$	 & 2.944 & 0.009 \\
$rms_{RV1}$ (\ms)& \multicolumn{2}{c}{45}\\
$rms_{RV2}$ (\ms)& \multicolumn{2}{c}{52}\\
$rms_{LC}$ (mmag)& \multicolumn{2}{c}{0.61}\\
\tableline
\end{tabular}
\end{center}
\end{table}
\end{document}